\documentclass[prb,onecolumn,showpacs,preprintnumbers,amsmath,amssymb,floatfix]{revtex4}

\usepackage{graphicx}

\begin{document}

\title{A variance-minimization scheme for optimizing Jastrow factors}

\author{N.~D.~Drummond and R.~J.~Needs}

\affiliation{TCM Group, Cavendish Laboratory, University of Cambridge,
Madingley Road, Cambridge CB3 0HE, United Kingdom}

\date{\today}

\begin{abstract}

We describe a new scheme for optimizing many-electron trial wave
functions by minimizing the unreweighted variance of the energy using
stochastic integration and correlated-sampling techniques.  The scheme
is restricted to parameters that are linear in the exponent of a
Jastrow correlation factor, which are the most important parameters in
the wave functions we use.  The scheme is highly efficient and allows
us to investigate the parameter space more closely than has been
possible before.  We search for multiple minima of the variance in the
parameter space and compare the wave functions obtained using
reweighted and unreweighted variance minimization.

\end{abstract}


\maketitle

\section{Introduction}

Accurate many-body wave functions are essential to the variational and
diffusion quantum Monte Carlo (VMC and DMC) methods, as the wave
function controls both the statistical efficiency and the accuracy of
these techniques.\cite{foulkes_2001} Optimizing many-body wave
functions is perhaps the most important technical issue facing
practitioners of these quantum Monte Carlo (QMC) techniques today, and
it consumes large quantities of human and computing resources.

Wave-function optimization schemes have usually involved minimizing
either the variational energy or its variance.  Although it is
generally believed that wave functions corresponding to the minimum
energy have more desirable properties, variance minimization has been
very widely used because it has proved easier to design robust
minimization techniques for this
purpose.\cite{umrigar_1988a,kent_1999}  The scheme introduced in this
article involves minimizing the \textit{unreweighted} variance.  We
describe a new method for evaluating this quantity, which greatly
accelerates the optimization of parameters that occur in a linear
fashion in the exponent of a Jastrow factor.  These are, in general,
the most important parameters in QMC trial wave functions.  The
optimization step does not involve a sum over electron configurations,
which means that we can use very large numbers of configurations.  The
unreweighted variance is in fact a quartic function of the linear
parameters in the Jastrow factor, and the minima of multidimensional
quartic functions can be located very rapidly.  The efficiency of our
scheme has enabled us to explore the minimization procedure and the
parameter space in detail, and to investigate the possible existence
of multiple minima.

The distinction between the reweighted or true variance and the
unreweighted variance is explained in Sec.~\ref{section:variance}.  In
Sec.~\ref{section:accelmethod} we describe our accelerated scheme for
calculating the unreweighted variance.  In
Sec.~\ref{section:unrew_var_nature} we use our new method to study the
unreweighted variance in parameter space.  The minima of the
reweighted and unreweighted variance need not coincide, and in
Sec.~\ref{section:var_E_minima} we investigate which minimum
corresponds to the lower energy.  We discuss the sampling of
configuration space and the flexibility of the trial wave function in
Secs.~\ref{section:sampling} and \ref{section:flex}.  In
Secs.~\ref{section:newopt_scaling} and \ref{section:newopt_timing} we
compare the efficiency of the ``standard'' and accelerated
variance-minimization methods, both in theory and practice.  Finally,
we draw our conclusions in Sec.~\ref{section:newopt_conclusions}.

Hartree atomic units (a.u.)\ are used throughout, in which the Dirac
constant, the magnitude of the electronic charge, the electronic mass,
and $4\pi$ times the permittivity of free space are unity: $\hbar =
|e| = m_e = 4 \pi \epsilon_0 = 1$.  All of our QMC calculations were
carried out using the \textsc{casino} package.\cite{casino}

\section{The energy and its variance}
\label{section:variance}

Consider a real trial wave function $\Psi({\bf R})$, where ${\bf R}$
is a point in the electron configuration space.  In VMC the energy is
written as
\begin{equation}
E = \frac{\int \Psi({\bf R})^2 E_L({\bf R}) \, d{\bf R}}{\int
\Psi({\bf R})^2 \, d{\bf R}},
\label{eqn:E_vmc}
\end{equation}
where the \textit{local energy}, $E_L$, is
\begin{equation}
E_L({\bf R}) = \Psi({\bf R})^{-1} \hat{H}({\bf R}) \Psi({\bf R}),
\label{eqn:loc_E_defn}
\end{equation}
and $\hat{H}$ is the Hamiltonian.  The variance of the energy is
\begin{equation}
\sigma^2 = \frac{\int \Psi({\bf R})^2 (E_L({\bf R}) - E)^2 \, d{\bf
R}}{\int \Psi({\bf R})^2 \, d{\bf R}}.
\label{eqn:sigma^2_vmc}
\end{equation}

We write the trial wave function as $\Psi^{\{ \alpha \}}({\bf R})$, to
denote that it depends on a set of free parameters, $\{ \alpha \}$.
Throughout this article, we confine our attention to the optimization
of parameters in the Jastrow factor.  The nodal surface of the trial
wave function is independent of such parameters.  Consider a set of
$N_C$ configurations $\{ {\bf R} \}$ distributed according to $\left(
\Psi^{\{ \alpha_0 \}}({\bf R}) \right)^2$ for some fixed parameter set
$\{ \alpha_0 \}$.  The variance $\sigma^2$ is then estimated for any
given parameter set $\{ \alpha \}$ using a correlated-sampling
procedure, which gives rise to the \textit{reweighted variance},
\begin{equation}
\sigma_w^2 = \frac{T_{\{ \alpha_0 \}}^{\{ \alpha \}}}{\left( T_{\{
\alpha_0 \}}^{\{ \alpha \}}\right)^2-\sum_{\bf R} \left( W_{\{
\alpha_0 \}}^{\{ \alpha \}}({\bf R}) \right)^2} \sum_{\bf R} \left(
E_L^{\{ \alpha \}}({\bf R}) - \bar{E_w} \right)^2 W_{\{ \alpha_0
\}}^{\{ \alpha \}}({\bf R}), \label{eqn:rew_var_def}
\end{equation}
where the \textit{reweighted energy} is
\begin{equation}
\bar{E}_w = \frac{1}{T_{\{ \alpha_0 \}}^{\{ \alpha \}}} \sum_{\bf R}
E_L^{\{ \alpha \}}({\bf R}) W_{\{ \alpha_0 \}}^{\{ \alpha \}}({\bf R}),
\end{equation}
which is an estimate of $E$, and the total weight is
\begin{equation}
T_{\{ \alpha_0 \}}^{\{ \alpha \}} = \sum_{\bf R}W_{\{ \alpha_0 \}}^{\{
\alpha \}}({\bf R}), \end{equation} and the weights $W$ are
\begin{equation} W_{\{ \alpha_0 \}}^{\{ \alpha \}}({\bf R}) =
\left( \frac{\Psi^{\{ \alpha \}} ({\bf R}) }{ \Psi^{\{ \alpha_0 \}}
({\bf R})} \right)^2.
\end{equation}

The \textit{unreweighted variance} as a function of parameter set $\{
\alpha \}$ is defined to be
\begin{equation}
\sigma_u^2 = \frac{1}{N_C-1} \sum_{\bf R} \left( E_L^{\{ \alpha
\}}({\bf R}) - \bar{E_u} \right)^2,
\end{equation}
where the \textit{unreweighted energy} is
\begin{equation}
\bar{E_u} = \frac{1}{N_C} \sum_{\bf R} E_L^{\{ \alpha \}}({\bf R}).
\end{equation}
The reweighted and unreweighted variances are identical when the same
set of configurations is used and ${\{ \alpha \}} = {\{ \alpha_0 \}}$.
However, for any given ${\{ \alpha_0 \}}$ they are different functions
of ${\{ \alpha \}}$, and there is no reason to expect that their
minima coincide with each other, or that either minimum should
coincide with that of the (reweighted) energy.

Both $\sigma_w^2$ and $\sigma_u^2$ are non-negative, but are zero when
$\Psi^{\{\alpha\}}$ is an eigenstate of $\hat{H}$.  The reweighted and
unreweighted variances are therefore reasonable cost functions for
wave-function optimizations.  The reweighted energy is also a
reasonable cost function.  However, the problem with the reweighted
energy and variance is that the weights $W$ may vary rapidly as the
parameters change, especially for large systems, which leads to
instabilities in optimization procedures.\cite{kent_1999}  It can be
shown that the wave function used to generate the configuration set
corresponds to a stationary point of $\bar{E_u}$ (for perfect
sampling).  In what follows we will mainly be interested in optimizing
linear parameters in the Jastrow factor, and in this case we have
proved that the wave function used to generate the configuration set
corresponds to the global maximum of the unreweighted
energy\cite{footnote_unreweighted_energy}. The unreweighted energy is
clearly not a suitable cost function.  From these considerations we
conclude that the cost function with the most suitable mathematical
properties for the stable optimization of wave functions within the
correlated-sampling approach is the unreweighted variance.

The usual variance-minimization procedure is to generate a set of
electron configurations $\{ {\bf R} \}$ distributed according to
$\left( \Psi^{\{ \alpha_0 \}}({\bf R}) \right)^2$ using VMC, and then
to minimize the reweighted or unreweighted energy variance over this
set.  Since the variance landscape depends on the distribution of
configurations, several \textit{cycles} of configuration generation
and optimization are normally carried out, with the optimized wave
function from the previous cycle being used in each VMC
configuration-generation phase.  We usually iterate several times and
choose the wave function that gives the lowest \textit{variational
energy}.  In the limit of perfect sampling, the reweighted variance is
equal to the actual variance, and is therefore independent of the
configuration distribution, so that the optimized parameters would not
change over successive cycles of reweighted variance minimization.
This is not the case for unreweighted variance minimization;
nevertheless, by carrying out a number of cycles, a
``self-consistent'' parameter set may be obtained.

\section{Accelerated evaluation of the unreweighted variance
  \label{section:accelmethod}}

\subsection{The Slater-Jastrow wave function}

Let $\Psi$ be a Slater-Jastrow wave function for a many-body system:
\begin{equation}
\Psi({\bf R}) = \exp[J({\bf R})] S({\bf R}),
\label{eqn:slater-jastrow}
\end{equation}
where $\exp[J]$ is the Jastrow factor, which contains free parameters
to be determined by an optimization method, and $S$ is the Slater wave
function, which may be an expansion in several determinants of
single-particle orbitals.

Suppose that $J$ contains linear parameters
$\alpha_1,\ldots,\alpha_P$, that is,
\begin{equation}
J({\bf R}) = \sum_{i=1}^P f_i({\bf R}) \alpha_i + J_0({\bf R}),
\label{eqn:linear_Jastrow_defn}
\end{equation} where
$f_1,\ldots,f_P$ and $J_0$ are known functions of ${\bf R}$, which
depend upon the particular form of Jastrow factor used and do not
contain any free parameters.  We use the form of Jastrow factor
described in detail in Ref.~\cite{ndd_jastrow}, which contains linear
parameters. However, some of the terms have a finite extent in space
and the associated cutoff lengths must appear nonlinearly in the
Jastrow factor.  These cutoff lengths can be set on physical grounds
or optimized using small numbers of parameters and configurations and
the standard variance-minimization procedure, but their values cannot
be obtained using the accelerated scheme described here.

\subsection{Derivation of the quartic polynomial}

The local energy for the Slater-Jastrow wave function of
Eq.~(\ref{eqn:slater-jastrow}) is
\begin{equation}
E_L({\bf R}) = - \frac{1}{2} \sum_{i=1}^P \sum_{j=1}^P
g_{ij}^{(2)}({\bf R}) \alpha_i \alpha_j - \frac{1}{2} \sum_{i=1}^P
g_i^{(1)}({\bf R}) \alpha_i - \frac{1}{2} g^{(0)}({\bf R}) +V({\bf R}),
\label{eqn:quad_E_L}
\end{equation}
where $V$ is the potential energy and
\begin{eqnarray}
g_{ij}^{(2)}({\bf R}) & = & \nabla f_i \cdot \nabla f_j \\
g_i^{(1)}({\bf R}) & = & 2 \nabla f_i \cdot \nabla J_0 + \nabla^2 f_i
+ 2 \frac{\nabla S}{S} \cdot \nabla f_i \\ g^{(0)}({\bf R}) & = &
|\nabla J_0|^2 + \nabla^2 J_0 + 2\frac{\nabla S}{S} \cdot \nabla J_0 +
\frac{\nabla^2 S}{S},
\label{eqn:gij_defn}
\end{eqnarray}
and we note that $g_{ij}^{(2)}=g_{ji}^{(2)}$.  The square of the local
energy is given by
\begin{eqnarray}
E_L^2({\bf R}) & = & \sum_{i=1}^P \sum_{j=1}^P \sum_{k=1}^P
\sum_{l=1}^P G_{ijkl}^{(4)}({\bf R}) \alpha_i \alpha_j \alpha_k
\alpha_l + \sum_{i=1}^P \sum_{j=1}^P \sum_{k=1}^P G_{ijk}^{(3)}({\bf
R}) \alpha_i \alpha_j \alpha_k \nonumber \\ & & {} + \sum_{i=1}^P
\sum_{j=1}^P G_{ij}^{(2)}({\bf R}) \alpha_i \alpha_j + \sum_{i=1}^P
G_i^{(1)}({\bf R}) \alpha_i +G^{(0)}({\bf R}),
\end{eqnarray}
where
\begin{eqnarray}
G_{ijkl}^{(4)}({\bf R}) & = & \frac{g_{ij}^{(2)}({\bf R})
g_{kl}^{(2)}({\bf R})}{4} \\ G_{ijk}^{(3)}({\bf R}) & = &
\frac{g_{ij}^{(2)}({\bf R}) g_k^{(1)}({\bf R})}{2} \\
G_{ij}^{(2)}({\bf R}) & = & \frac{g_i^{(1)}({\bf R}) g_j^{(1)}({\bf
R})}{4} - g_{ij}^{(2)}({\bf R}) \left( V({\bf R})-\frac{g^{(0)}({\bf
R})}{2} \right) \\ G_i^{(1)}({\bf R}) & = & -g_i^{(1)}({\bf R}) \left(
V({\bf R})-\frac{g^{(0)}({\bf R})}{2} \right) \\ G^{(0)}({\bf R}) & =
& \left( V({\bf R})-\frac{g^{(0)}({\bf R})}{2} \right)^2.
\end{eqnarray}
(Note that
$G_{ijkl}^{(4)}=G_{jikl}^{(4)}=G_{ijlk}^{(4)}=G_{klij}^{(4)}$,
$G_{ijk}^{(3)}=G_{jik}^{(3)}$, and $G_{ij}^{(2)}=G_{ji}^{(2)}$.)

Suppose the VMC method is used to generate a set of $N_C$ points in
configuration space, $\{ {\bf R} \}$, which are distributed according
to the square of an approximate trial wave function.  For any quantity
$A({\bf R})$, let
\begin{equation}
\bar{A}=\frac{1}{N_C} \sum_{\bf R} A({\bf R})
\end{equation}
be the average of $A({\bf R})$ over the set of $N_C$ configurations.
The unreweighted variance may be written as
\begin{eqnarray}
\sigma_u^2 & = & \frac{N_C}{N_C-1} \left( \bar{E_L^2} - \bar{E_L}^2
\right) \nonumber \\ & \equiv & \frac{N_C}{N_C-1} \left( \sum_{i=1}^P
\sum_{j=1}^P \sum_{k=1}^P \sum_{l=1}^P K_{ijkl}^{(4)} \alpha_i
\alpha_j \alpha_k \alpha_l + \sum_{i=1}^P \sum_{j=1}^P \sum_{k=1}^P
K_{ijk}^{(3)} \alpha_i \alpha_j \alpha_k \right. \nonumber \\ & & {}
\left. + \sum_{i=1}^P \sum_{j=1}^P K_{ij}^{(2)} \alpha_i \alpha_j +
\sum_{i=1}^P K_i^{(1)} \alpha_i + K^{(0)} \right),
\label{eqn:lsq_in_terms_of_K}
\end{eqnarray}
where
\begin{eqnarray}
K_{ijkl}^{(4)} & = & \bar{G}_{ijkl}^{(4)} - \frac{\bar{g}_{ij}^{(2)}
\bar{g}_{kl}^{(2)}}{4} \\ K_{ijk}^{(3)} & = & \bar{G}_{ijk}^{(3)} -
\frac{\bar{g}_{ij}^{(2)} \bar{g}_k^{(1)}}{2} \\ K_{ij}^{(2)} & = &
\bar{G}_{ij}^{(2)} - \frac{\bar{g}_i^{(1)} \bar{g}_j^{(1)}}{4} +
\bar{g}_{ij}^{(2)} \left( \bar{V}-\frac{\bar{g}^{(0)}}{2} \right) \\
K_i^{(1)} & = & \bar{G}_i^{(1)} + \bar{g}_i^{(1)} \left(
\bar{V}-\frac{\bar{g}^{(0)}}{2} \right) \\ K^{(0)} & = &
\bar{G}^{(0)}-\left( \bar{V} - \frac{\bar{g}^{(0)}}{2} \right)^2.
\end{eqnarray}
(Note that
$K_{ijkl}^{(4)}=K_{jikl}^{(4)}=K_{ijlk}^{(4)}=K_{klij}^{(4)}$,
$K_{ijk}^{(3)}=K_{jik}^{(3)}$, and $K_{ij}^{(2)}=K_{ji}^{(2)}$.)  The
unreweighted variance is quartic in the set of free parameters.  Once
the values of $K^{(n)}$ have been computed, there is no need to
perform any further summations over the set of configurations during
the optimization of the parameters.

Throughout this article the potential energy is assumed to be a local
operator, so the local potential energy is independent of the
wave-function parameters.  When the variance-minimization algorithm is
applied to systems containing pseudoatoms, the change in the local
potential energy due to the nonlocal part of the pseudopotential is
neglected. Not only does this greatly improve the speed of the
variance-minimization process, but it also appears to improve the
stability of the algorithm.

\subsection{Evaluating the least-squares function during an optimization}

\subsubsection{Accumulating $\bar{G}$ and $\bar{g}$}

The values of $\bar{g}$, $\bar{G}$, and $\bar{V}$ are accumulated
during a VMC simulation by keeping a running total of the values of
$g({\bf R})$, $G({\bf R})$, and $V({\bf R})$ encountered at each step
of the random walk; there is no need to store data for each
configuration.  The accumulated elements of $G$ are stored in a
one-dimensional array and, furthermore, the symmetries of $G$ are
exploited in order to minimize the length of this vector.  The numbers
of $G^{(4)}$, $G^{(3)}$, $G^{(2)}$, $G^{(1)}$, and $G^{(0)}$ elements
to be calculated and stored are
\begin{eqnarray}
N_G^{(4)} & = & \frac{\left( \frac{P(P+1)}{2} \right)
\left(\frac{P(P+1)}{2}+1 \right)}{2}
\label{eqn:NG4_exp} \\ N_G^{(3)} & = &  \frac{P^2(P+1)}{2} \\ N_G^{(2)} & = &
\frac{P(P+1)}{2} \\ N_G^{(1)} & = & P \\ N_G^{(0)} & = & 1,
\end{eqnarray}
respectively.   $G_{ijkl}^{(4)}$ is symmetric with respect to $i$ and
$j$, and is also symmetric with respect to $k$ and $l$.  In order to
label the independent elements of $G^{(4)}$, one can replace $(i,j)$
by a single index $I$ that takes $P(P+1)/2$ different values.
Likewise, $(k,l)$ can be replaced by a single index $J$ that takes
$P(P+1)/2$ different values. $G^{(4)}_{IJ}$ is still symmetric with
respect to $I$ and $J$; hence $(I,J)$ can be replaced by a single
index $K$ which takes $N_G^{(4)}$ different values, where $N_G^{(4)}$
is given in Eq.~(\ref{eqn:NG4_exp}).  This is the method by which the
elements of $G^{(4)}$ are indexed in practice.  Counting and indexing
the elements of $G^{(3)}$, $G^{(2)}$, and $G^{(1)}$ are relatively
straightforward.  The total number of $G$ elements grows as $O(P^4)$.
Storing these coefficients represents the memory bottleneck for the
accelerated optimization procedure.  With $P=30$ parameters (a typical
number), 122,791 $G$ elements must be stored.  With $P=100$ parameters
(a large number), 13,263,926 elements must be stored.  Alternatively,
the number of elements to be \textit{stored} could be reduced by using
the same strategy as that suggested in Sec.~\ref{section:reduced_LSF}
for evaluating the unreweighted variance.  This would not affect the
number of elements that have to be \textit{evaluated}, however, and it
may slow down the VMC calculation even further.  The saving in memory
would typically be a factor of between 2.5 and 3, which is
insignificant, given the $O(P^4)$ scaling of the method.

\subsubsection{Evaluating the least-squares function
  \label{section:reduced_LSF}}

Before the start of the optimization, the coefficient of each
different product of parameters is computed and the coefficients are
stored in a one-dimensional array.  This allows the unreweighted
variance to be evaluated extremely rapidly.  The set of possible
products of four of the parameters is $\{ \alpha_i \alpha_j \alpha_k
\alpha_l ~:~ i \leq j \leq k \leq l \}$, and similarly for the
products of three and two parameters.  So the unreweighted variance
can be written as
\begin{equation}
\sigma_u^2 = \frac{N_C}{N_C-1} \left( \Gamma^{(0)}+\sum_{i=1}^P
\alpha_i \left( \Gamma_i^{(1)}+ \sum_{j=i}^P \alpha_j \left(
\Gamma_{ij}^{(2)} + \sum_{k=j}^P \alpha_k \left(
\Gamma_{ijk}^{(3)}+\sum_{l=k}^P \alpha_l \Gamma_{ijkl}^{(4)} \right)
\right) \right) \right),
\label{eqn:lsq_in_terms_of_Gamma}
\end{equation}
where the $\Gamma^{(n)}$ are defined in terms of $K^{(n)}$ (see below)
and are stored as one-dimensional arrays. The number of elements of
$\Gamma^{(n)}$ is given by the number of distinct products of $n$
parameters, which can be shown to be
\begin{equation}
N_T^{(n)} = {P+n-1 \choose n},
\label{eqn:no_terms_Gamma}
\end{equation}
while the total number of elements of the $\Gamma$ arrays is
\begin{equation}
N_T = {P+4 \choose 4},
\end{equation}
which increases as $O(P^4)$.  For $P=30$ parameters, the number of
terms that must be summed over to obtain the unreweighted variance is
46,376, while for $P=100$ parameters, the number of terms is 4,598,126.

For each $\{i,j,k,l\}$ with $i \leq j \leq k \leq l$,
$\Gamma^{(4)}_{ijkl}$ is equal to the sum of $K^{(4)}_{ijkl}$ over all
distinct permutations of $\{i,j,k,l\}$.  $\Gamma^{(3)}$,
$\Gamma^{(2)}$, $\Gamma^{(1)}$, and $\Gamma^{(0)}$ are constructed in
a similar fashion.

\subsubsection{Derivatives of the least-squares function
\label{section:grad_LSF}}

Derivatives of the unreweighted variance are given by
\begin{equation}
\frac{\partial \sigma_u^2}{\partial \alpha_n} = \frac{N_C}{N_C-1}
\left( \sum_{i=1}^P \sum_{j=1}^P \sum_{k=1}^P M^{(3)}_{ijk}(n)
\alpha_i \alpha_j \alpha_k + \sum_{i=1}^P \sum_{j=1}^P M^{(2)}_{ij}(n)
\alpha_i \alpha_j + \sum_{i=1}^P M^{(1)}_i(n) \alpha_i + M^{(0)}(n)
\right),
\label{eqn:lsf_derivs}
\end{equation}
where
\begin{eqnarray}
M^{(3)}_{ijk}(n) & = &
K_{nijk}^{(4)}+K_{injk}^{(4)}+K_{ijnk}^{(4)}+K_{ijkn}^{(4)} \\
M^{(2)}_{ij}(n) & = & K_{nij}^{(3)}+K_{inj}^{(3)}+K_{ijn}^{(3)} \\
M^{(1)}_i(n) & = & K_{ni}^{(2)}+K_{in}^{(2)} \\ M^{(0)}(n) & = &
K^{(1)}_n.
\end{eqnarray}
In practice, derivatives are evaluated as
\begin{equation}
\frac{\partial \sigma_u^2}{\partial \alpha_n} = \frac{N_C}{N_C-1}
\left( \Lambda^{(0)}(n)+\sum_{i=1}^P \alpha_i \left(
\Lambda_i^{(1)}(n)+\sum_{j=i}^P \alpha_j \left( \Lambda_{ij}^{(2)}(n)
+ \sum_{k=j}^P \alpha_k \Lambda_{ijk}^{(3)}(n) \right) \right) \right),
\end{equation}
where the $\Lambda(n)$ are defined in terms of the $M(n)$ in an
analogous fashion to the definition of $\Gamma$ in terms of $K$ in
Sec.~\ref{section:reduced_LSF}.  The total number of elements of
$\Lambda$ is
\begin{equation}
N_T^\prime = P {P+3 \choose 3},
\end{equation}
which grows as $O(P^4)$.  The $\Lambda$ arrays used to evaluate the
gradient of the unreweighted variance may be somewhat larger than the
$G$ arrays.

\subsection{Minimizing the variance \label{section:opt_methods}}

Ideally, one would like to use an optimization method that enables one
to find the \textit{global} minimum of the variance with respect to
the wave-function parameters.  Unfortunately, existing
variance-minimization algorithms generally use numerical optimization
methods which, if started close to a particular local minimum, will
always converge to that minimum.  However, in the case of the quartic
unreweighted variance in the space of linear Jastrow parameters, it is
relatively easy to carry out an extensive search for the global
minimum.

Standard methods for minimizing a function of many variables include
the method of steepest descents, the conjugate-gradients method, and
the BFGS method.\cite{press_F77} Of these three methods, we have found
the BFGS algorithm to converge most rapidly for a wide variety of test
systems.

Along any given line in the space of linear Jastrow parameters the
unreweighted variance is a quartic polynomial of a single variable.
The method by which the variance along a line can be re-expressed as a
quartic polynomial is given in Appendix
\ref{section:line_min_details}.  A quartic polynomial of a single
variable has at most two minima on the real axis.  The gradient of a
quartic function is a cubic, whose three roots can be obtained
analytically;\cite{press_F77} hence it is straightforward to locate
the global minimum of the unreweighted variance along the line.

In order to search the parameter space for the global minimum of the
variance with respect to the linear Jastrow parameters, we first
perform a BFGS minimization.  Starting from this minimum we choose
directions at random and use the analytic line-minimization technique
to search for a second minimum, lower than the first.  If a second
minimum is found then BFGS is used to converge to the new minimum, and
the process is repeated.

\section{The nature of the unreweighted variance
  \label{section:unrew_var_nature}}

\subsection{Linear Jastrow parameters \label{section:lsf_nature_lin_param}}

We used the method described in Sec.~\ref{section:opt_methods} to
search for minima when optimizing the linear Jastrow parameters in the
SiH$_4$ molecule, the all-electron neon atom, a 16-atom cell of
diamond-structure pseudosilicon subject to periodic boundary
conditions, and an electron-hole gas.  However, even sampling up to
$10^7$ random directions, multiple minima were \textit{only} found
when the configuration space was deliberately sampled extremely poorly.

\begin{figure}
\begin{center}
\includegraphics*{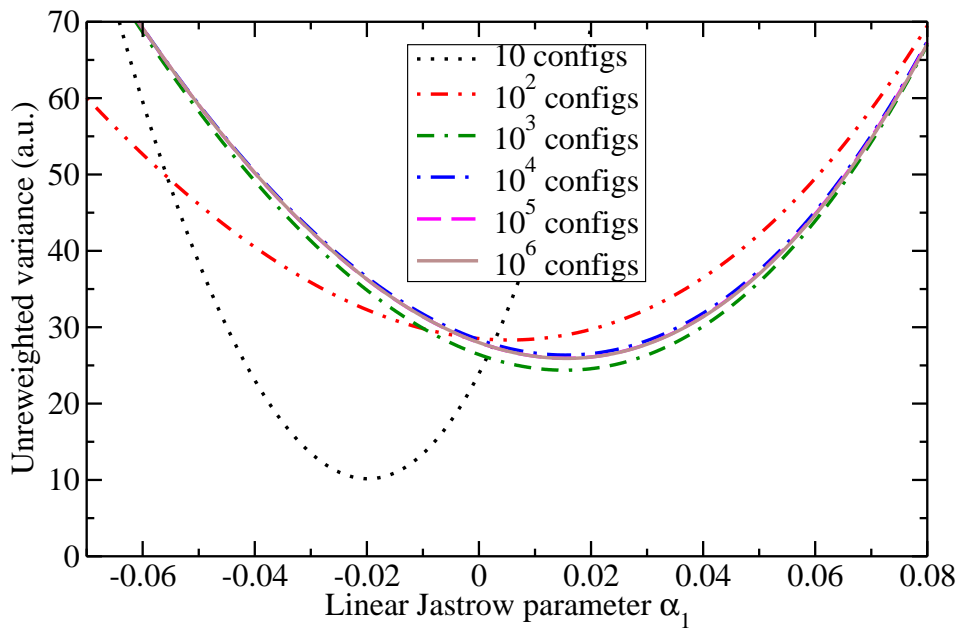}
\caption{The unreweighted variance $\sigma_u^2$ for an all-electron
  neon atom plotted against one of the linear Jastrow parameters.  All
  of the other linear parameters are set to zero.  Different numbers
  of configurations were used to calculate the quartic coefficients of
  the unreweighted variance.  In each case the configurations were
  distributed according to the square of the Hartree-Fock wave
  function.  The Jastrow factor contained a total of 24 linear
  parameters.  The curves for $10^5$ and $10^6$ configurations are
  indistinguishable in the figure.
  \label{figure:neon_LSF_v_alpha1_P24}}
\end{center}
\end{figure}

Plots of the unreweighted variance against the value of one of the
linear parameters are shown for an all-electron neon atom in
Fig.~\ref{figure:neon_LSF_v_alpha1_P24}.  It can be seen that the
unreweighted variance converges to a limit as  the number of
configurations is increased.  There is only one minimum in every case.

\begin{figure}
\begin{center}
\includegraphics*{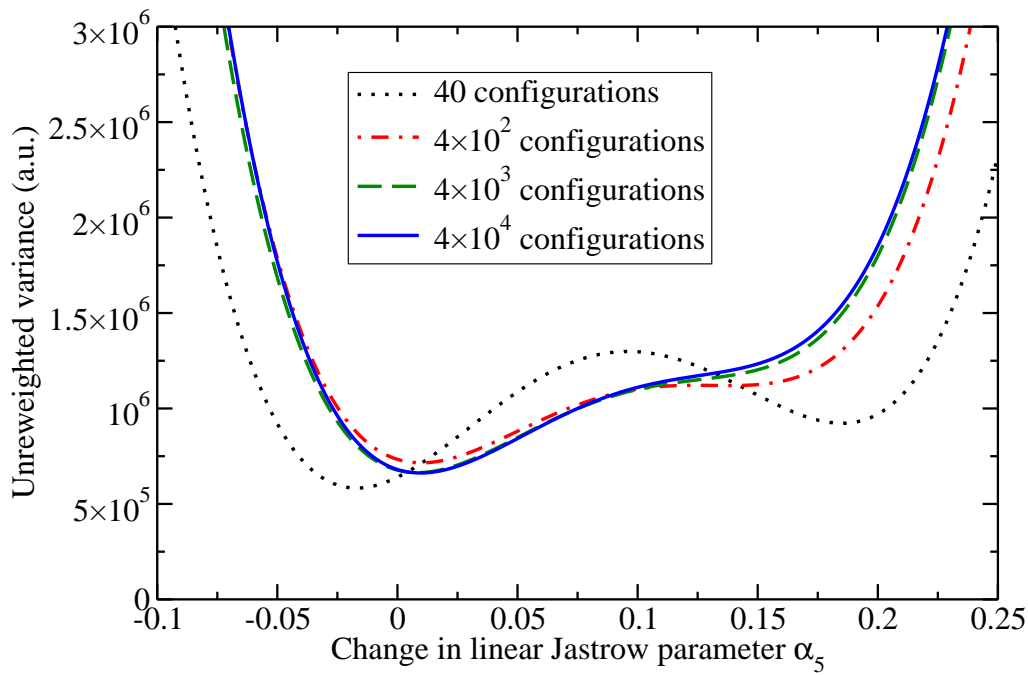}
\caption{The unreweighted variance $\sigma_u^2$ for an all-electron
  neon atom plotted against the change in one of the linear Jastrow
  parameters.  All the parameters in the Jastrow parameter are set to
  large, random values.  Different numbers of VMC-generated
  configurations were used to calculate the quartic coefficients of
  the unreweighted variance.  In each case the configurations were
  distributed according to the square of a Slater-Jastrow wave
  function, where the Jastrow factor contained the random parameters,
  so that the resulting distribution was very unlike the ground-state
  distribution.  The Jastrow factor contained a total of 72 linear
  parameters.  The Slater wave function contained Hartree-Fock
  orbitals.
  \label{figure:LSF_v_alpha5_neon}}
\end{center}
\end{figure}

Plots of the unreweighted variance for an all-electron neon atom
against the value of a parameter for an extremely poor sampling of
configuration space are shown in Fig.~\ref{figure:LSF_v_alpha5_neon}.
When few configurations were used ($N_C=40$), it was possible to find
two minima of the variance along lines in parameter space, proving
that nonglobal minima can exist.  However, it was also found that
increasing the number of configurations tended to prevent the
occurrence of two minima along lines in parameter space.

\subsection{Nonlinear Jastrow parameters}

\begin{figure}
\begin{center}
\includegraphics*{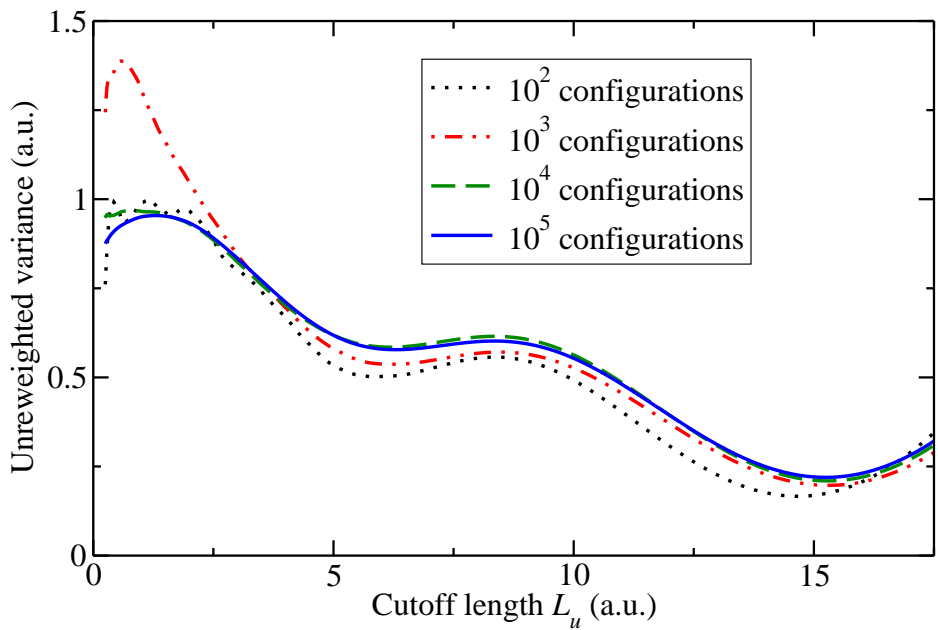}
\caption{The unreweighted variance $\sigma_u^2$ for an SiH$_4$
  molecule (with a Hartree-Fock silicon
  pseudopotential\cite{lee_thesis}) plotted against the cutoff length
  for the electron-electron terms in the Jastrow factor, $L_u$. The
  Jastrow factor is such that the local energy is continuous when an
  electron-electron separation passes through the cutoff
  length.\cite{ndd_jastrow}  Different numbers of VMC-generated
  configurations were used to calculate the unreweighted variance.
  All of the linear Jastrow parameters are set to zero.  In each case
  the configurations were distributed according to the square of the
  Hartree-Fock wave function.  The Jastrow factor contained a total of
  56 linear parameters, plus three cutoff
  lengths. \label{figure:LSF_v_Lu_silane_C3_zeroparams}}
\end{center}
\end{figure}

\begin{figure}
\begin{center}
\includegraphics*{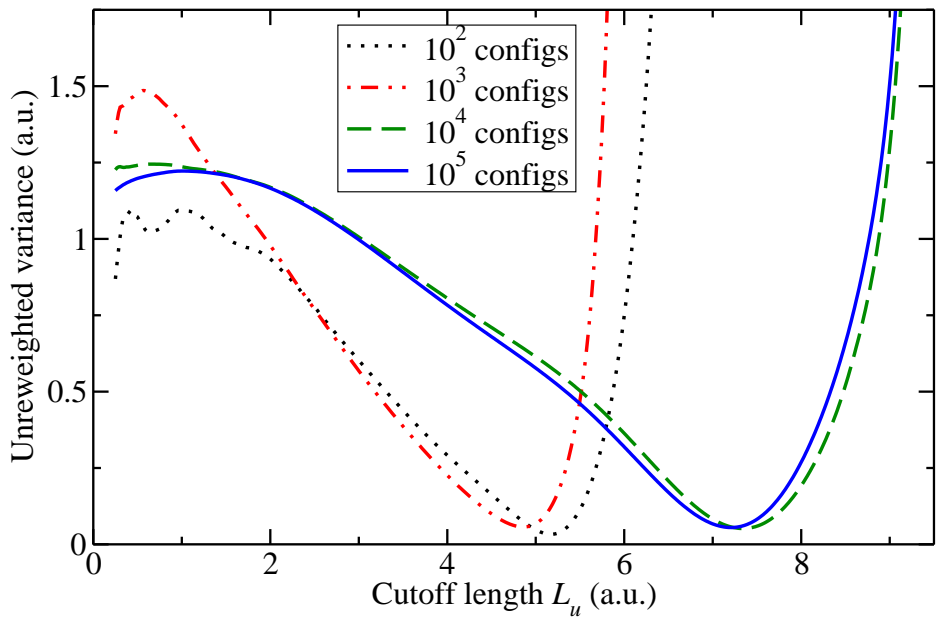}
\caption{This figure is the same as
  Fig.~\ref{figure:LSF_v_Lu_silane_C3_zeroparams}, except that all of
  the parameters in the Jastrow factor (including the cutoff lengths)
  have been optimized.  \label{figure:LSF_v_Lu_silane_C3_optparams}}
\end{center}
\end{figure}

Plots of the unreweighted variance of the SiH$_4$ molecule against a
nonlinear Jastrow parameter---the cutoff length for the
electron-electron correlation term\cite{ndd_jastrow}---are shown in
Figs.~\ref{figure:LSF_v_Lu_silane_C3_zeroparams} and
\ref{figure:LSF_v_Lu_silane_C3_optparams}.  The behavior of the
unreweighted variance is far worse when the cutoff length is varied
than when a linear parameter is varied: the variance has multiple
minima along lines in parameter space and there is some noise in the
variance, especially for poor samplings of configuration space.   It
can be seen in Fig.~\ref{figure:LSF_v_Lu_silane_C3_optparams} that the
optimized cutoff lengths obtained using $10^2$ or $10^3$
configurations are considerably shorter than the cutoff lengths
obtained using $10^4$ or $10^5$ configurations.  In the former case
the cutoff lengths are trapped in the nonglobal minimum that can be
seen in Fig.~\ref{figure:LSF_v_Lu_silane_C3_zeroparams}, while the
deeper minimum is reached in the latter case.  The Jastrow factor used
to produce Figs.~\ref{figure:LSF_v_Lu_silane_C3_zeroparams} and
\ref{figure:LSF_v_Lu_silane_C3_optparams} is such that the local
energy is continuous when an electron-electron separation passes
through the cutoff length.  If a Jastrow factor that gives rise to a
discontinuous local energy at the cutoff length were to be used, the
variance would be an extremely noisy function of the cutoff length,
especially for thin samplings of configuration space.  Optimization of
the cutoff lengths for such Jastrow factors has been found to be very
difficult.\cite{ndd_jastrow} The existence of multiple minima when
cutoff lengths are optimized suggests that it may be worthwhile
performing variance-minimization calculations using several different
initial cutoff lengths.

\section{Minima of the variance and the energy \label{section:var_E_minima}}

\subsection{The reweighted and unreweighted variance}

\subsubsection{Plots of the reweighted and unreweighted variance}

\begin{figure}
\begin{center}
\includegraphics*{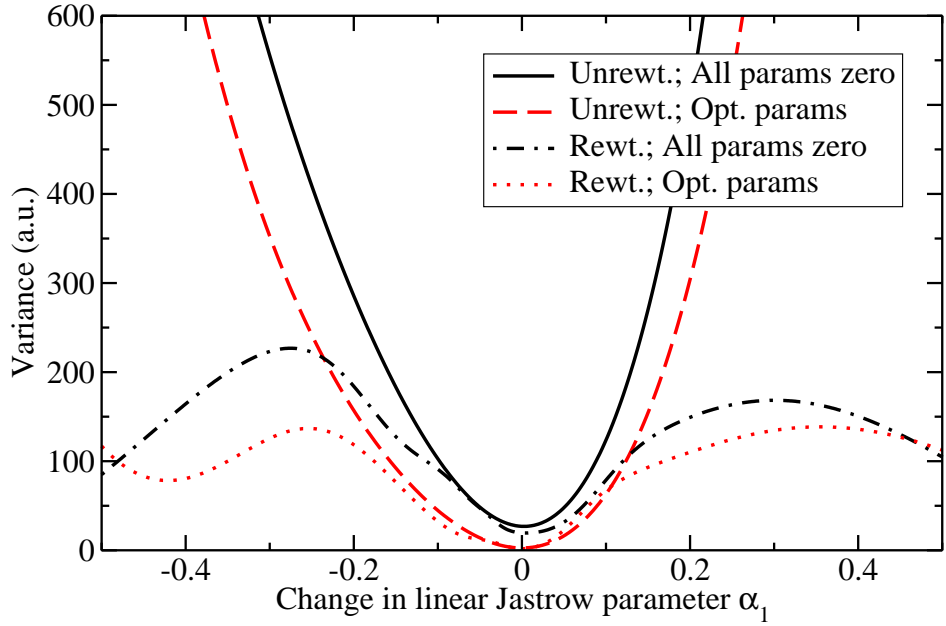}
\caption{The reweighted and unreweighted variance for an all-electron
  neon atom plotted against the change in the value of a linear
  Jastrow parameter, $\alpha_1$. Plots are shown for the case in which
  all the parameters are set to zero and the case in which all the
  parameters have been optimized.  The set of 100 configurations used
  to calculate the variance were distributed according to the square
  of the Hartree-Fock wave function.  The Jastrow factor contained a
  total of 27 linear parameters.
  \label{figure:LSF_v_alpha1_wt_unwt_Nc100_neon}}
\end{center}
\end{figure}

\begin{figure}
\begin{center}
\includegraphics*{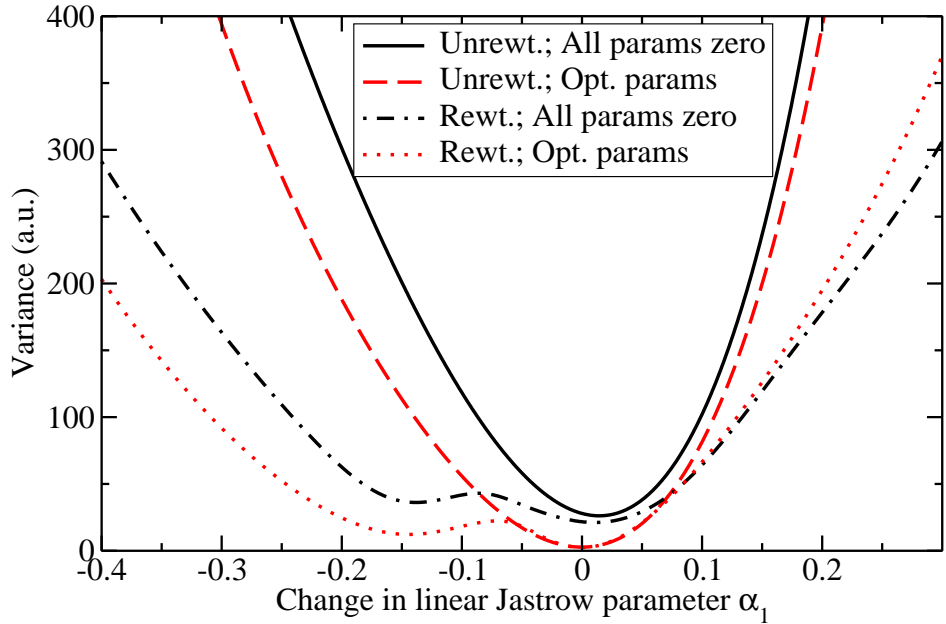}
\caption{This figure is the same as Figure
  \ref{figure:LSF_v_alpha1_wt_unwt_Nc100_neon} except that $10^4$
  configurations were used to calculate the
  variance. \label{figure:LSF_v_alpha1_wt_unwt_Nc10000_neon}}
\end{center}
\end{figure}

Plots of the reweighted and unreweighted variances for an all-electron
neon atom against one of the linear Jastrow parameters are shown in
Figs.~\ref{figure:LSF_v_alpha1_wt_unwt_Nc100_neon} and
\ref{figure:LSF_v_alpha1_wt_unwt_Nc10000_neon} for small and large
numbers of configurations.  The reweighted and unreweighted variances
have their minima in different places, with their values at the minima
being different from one another.  The variance is a smooth function
of the linear Jastrow parameter in each case, but there are multiple
minima of the reweighted variance along lines in parameter space,
demonstrating that nonglobal minima can exist.  Furthermore, the
minima of the reweighted variance are not as sharply defined as those
of the unreweighted variance.  Minimization of the unreweighted
variance is therefore more likely to be rapid and stable.

\subsubsection{Quality of variance-minimization results}

The outcomes of reweighted and unreweighted variance-minimization
calculations are shown in Table \ref{table:neon_varmin_NC5000}.  For a
relatively sparse sampling of configuration space, reweighted variance
minimization is pathologically unstable, while unreweighted variance
minimization is perfectly well-behaved.  For a dense sampling of
configuration space the two methods give very similar results, and
there is no evidence that the reweighted variance-minimization
algorithm performs any better than the unreweighted algorithm, or
\textit{vice versa}.

\begin{table}
\begin{center}
\begin{tabular}{lccr@{.}lr@{.}lr@{.}lr@{.}l}
\hline \hline

& & & \multicolumn{4}{c}{VMC energy (a.u.)} &
\multicolumn{4}{c}{Variance (a.u.)} \\

\raisebox{1em}[0pt]{$P$} & \raisebox{1em}[0pt]{$N_C$} &
\raisebox{1em}[0pt]{Cycle} & \multicolumn{2}{c}{Unrew.} &
\multicolumn{2}{c}{Rew.} & \multicolumn{2}{c}{Unrew.} &
\multicolumn{2}{c}{Rew.} \\

\hline

1 & 500 & 1 & $-128$&$5469(5)$ & $-128$&$5469(5)$ & $29$&$3(1)$ &
$29$&$3(1)$ \\

1 & 500 & 2 & $-128$&$5424(3)$ & $-128$&$5858(3)$ & $6$&$3786(7)$ &
$6$&$220(2)$ \\

1 & 500 & 3 & $-128$&$6178(3)$ & $-128$&$6129(3)$ & $6$&$310(1)$ &
$6$&$207(1)$ \\

1 & 500 & 4 & $-128$&$6267(3)$ & $-86$&$508(3)$ & $6$&$1564(7)$ &
$1501$&$1(2)$ \\

\hline

1 & $5 \times 10^5$ & 2 & $-128$&$6170(3)$ & $-128$&$6248(3)$ &
$6$&$116(2)$ & $6$&$0848(9)$ \\

1 & $5 \times 10^5$ & 3 & $-128$&$6226(3)$ & $-128$&$6261(3)$ &
$6$&$096(1)$ & $6$&$0844(8)$ \\

1 & $5 \times 10^5$ & 4 & $-128$&$6201(3)$ & $-128$&$6260(3)$ &
$6$&$103(1)$ & $6$&$0834(9)$ \\

\hline

72 & 500 & 2 & $-128$&$87184(9)$ & $-128$&$86664(9)$ & $1$&$4178(7)$ &
$1$&$3733(8)$ \\

72 & 500 & 3 & $-128$&$88040(9)$ & $-50$&$00(9)$ & $1$&$358(1)$ &
$420$&$7(86)$ \\

72 & 500 & 4 & $-128$&$8749(1)$ & $258$&$83(1)$ & $1$&$421(2)$ &
$13599$&$1(4816)$ \\

\hline

72 & $5 \times 10^5$ & 2 & $-128$&$89760(7)$ & $-128$&$89622(7)$ &
$1$&$136(1)$ & $1$&$1330(9)$ \\

72 & $5 \times 10^5$ & 3 & $-128$&$89742(7)$ & $-128$&$89677(7)$ &
$1$&$1343(7)$ & $1$&$132(1)$ \\

72 & $5 \times 10^5$ & 4 & $-128$&$89752(7)$ & $-128$&$89655(7)$ &
$1$&$137(2)$ & $1$&$1326(9)$ \\

\hline \hline
\end{tabular}
\caption{Results of reweighted and unreweighted variance-minimization
  calculations for an all-electron neon atom.  $P$ is the number of
  linear parameters in the Jastrow factor and $N_C$ is the number of
  configurations used to perform the optimization.  (Long VMC runs
  were used to obtain the energies and variances shown in the table.)
  Only linear Jastrow parameters were optimized.  The VMC energy and
  variance for cycle 1 are estimates of the Hartree-Fock energy and
  variance, and are the same for each $P$ and $N_C$.
  \label{table:neon_varmin_NC5000}}
\end{center}
\end{table}

\subsection{Coincidence of the minima of the energy and the variance
  \label{section:E_var_minima}}

\begin{figure}
\begin{center}
\includegraphics*{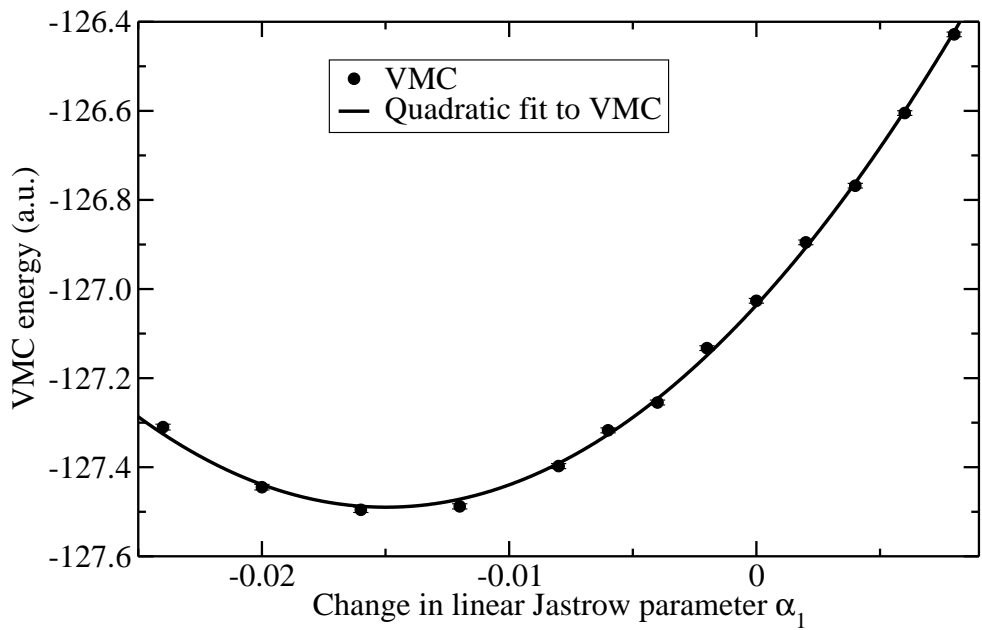}
\caption{The VMC energy against the change in the value of a linear
  Jastrow parameter $\alpha_1$ from the value determined by
  self-consistent unreweighted variance minimization.  The Jastrow
  factor was chosen to be poor, with no electron-nucleus or
  electron-electron-nucleus terms, and the same electron-electron
  terms were used for both parallel and antiparallel spins.  There is
  only one optimizable parameter: $\alpha_1$.
  \label{figure:energy_v_alpha1_neon_poorwfn}}
\end{center}
\end{figure}

\begin{figure}
\begin{center}
\includegraphics*{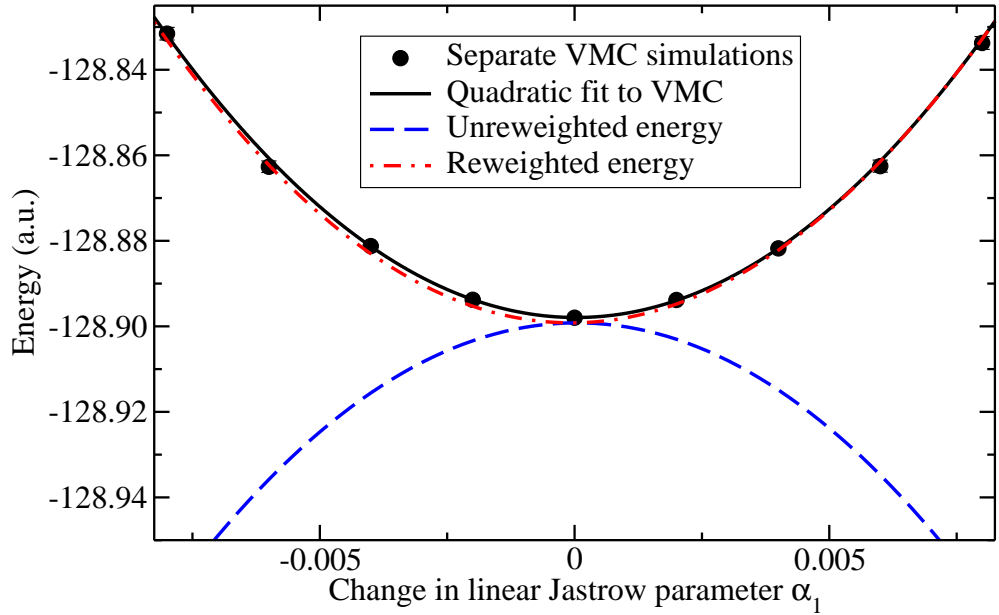}
\caption{The VMC energy of an all-electron neon atom against the
  change in the value of a linear Jastrow parameter $\alpha_1$ from
  the value determined by self-consistent unreweighted variance
  minimization.  Specifically, the Jastrow factor was the best
  all-electron neon Jastrow factor described in
  Ref.~\cite{ndd_jastrow}.  The reweighted and unreweighted energies
  calculated using $8 \times 10^5$ configurations distributed
  according to the square of the optimized wave function are also
  plotted. The statistical error bars in the VMC data are smaller than
  the symbols. \label{neon_VMC_E_v_alpha1}}
\end{center}
\end{figure}

As is clearly demonstrated in
Fig.~\ref{figure:energy_v_alpha1_neon_poorwfn}, the self-consistent
minimum of the unreweighted variance does not necessarily coincide
with the minimum of the VMC energy.  On the other hand, for a
high-quality Jastrow factor, the minima of the unreweighted variance
and energy are generally in close agreement, as is shown in
Fig.~\ref{neon_VMC_E_v_alpha1}.  We have no evidence, for all-electron
atoms at least, that any significant advantage could be obtained by
optimizing linear Jastrow parameters in a good Jastrow factor using an
energy-minimization method.  It can also be seen in
Fig.~\ref{neon_VMC_E_v_alpha1} that the reweighted energy follows the
actual VMC energy data closely (the statistical error in the
reweighted energy at the optimal wave function is $0.001$\,a.u.).
This implies that, provided enough configurations are used, the wave
function could be optimized by reweighted energy minimization.

\begin{figure}
\begin{center}
\includegraphics*{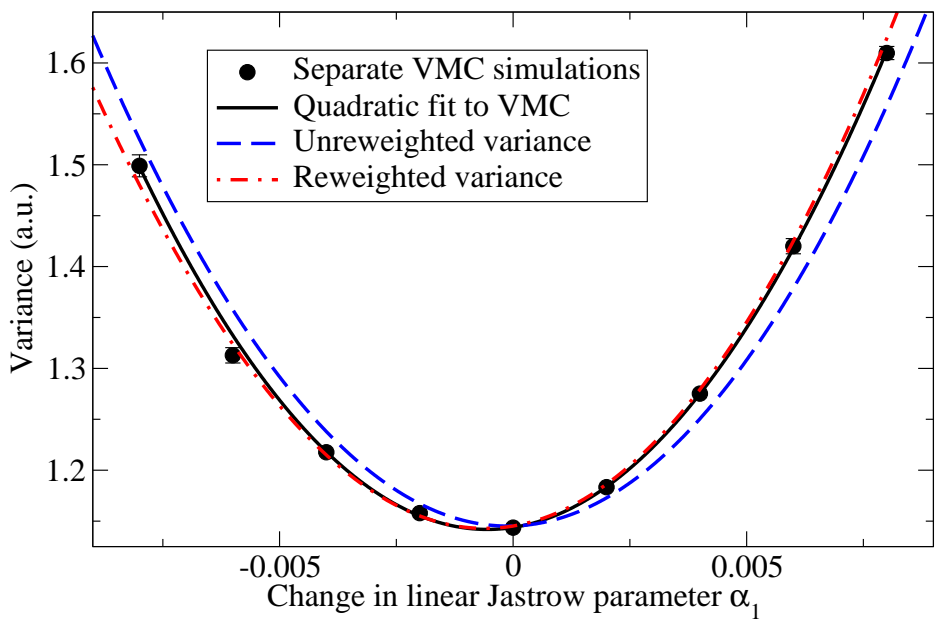}
\caption{The VMC variance for an all-electron neon atom against the
  change in the value of a linear Jastrow parameter $\alpha_1$ from
  its value determined by self-consistent unreweighted variance
  minimization.  Specifically, the Jastrow factor was the best
  all-electron neon Jastrow factor described in
  Ref.~\cite{ndd_jastrow}  The reweighted and unreweighted variances
  calculated using $8 \times 10^5$ configurations distributed
  according to the square of the optimized wave function are also
  plotted. Where the error bars in the VMC data cannot be seen, they
  are smaller than the symbols. \label{neon_VMC_var_v_alpha1}}
\end{center}
\end{figure}

A plot of the VMC energy variance against the change in a linear
Jastrow parameter from its optimal value in an all-electron neon atom
is shown in Fig.~\ref{neon_VMC_var_v_alpha1}.  As one would expect,
the reweighted variance matches the actual variance, unlike the
unreweighted variance; however, there is no significant difference
between the minima of the variance and the unreweighted variance (and
hence the energy).

We have also studied the question of the coincidence of the minima of
the energy, the variance, and the self-consistent unreweighted
variance using a variety of model systems, for which the integrals
could be performed exactly.  The models consisted of one-dimensional
potential wells and various trial wave functions with a single
variable parameter.  We studied several examples for a single particle
and an example for two identical, interacting fermions.  These
examples showed that the global minima of the energy, the variance and
the self-consistent unreweighted variance can be different.  In all
cases studied the parameters optimized by self-consistent unreweighted
variance minimization gave lower energies than the parameters
optimized by reweighted or ``true'' variance minimization.
Furthermore, in many cases, the parameters from the self-consistent
unreweighted variance minimum coincided exactly with the
energy-minimized parameters, suggesting that some underlying principle
was at work.

\section{The sampling of configuration space \label{section:sampling}}

\subsection{The number of configurations}

\begin{figure}
\begin{center}
\includegraphics*{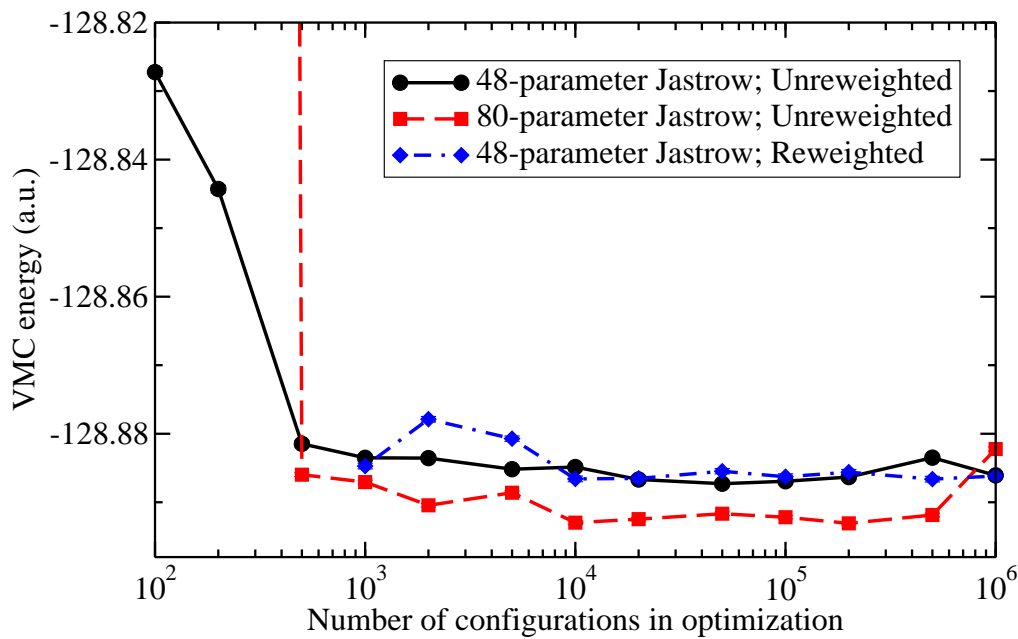}
\caption{The VMC energy of an all-electron neon atom against the
  number of configurations used to optimize the linear Jastrow
  parameters in an unreweighted variance minimization.  Eight
  optimization cycles were performed for each number of configurations
  in order to ensure that self-consistency was achieved.  The Slater
  wave function contained Hartree-Fock orbitals.  The error bars in
  the VMC data are smaller than the symbols.
  \label{VMC_E_v_Nc_neon}}
\end{center}
\end{figure}

The VMC energy for a neon-atom Slater-Jastrow wave function is plotted
against the number of configurations used to optimize the Jastrow
factor in Fig.~\ref{VMC_E_v_Nc_neon}.  It can be seen that the
wave-function quality improves very rapidly, then saturates at between
$5 \times 10^2$ and $10^4$ configurations, for both small and large
numbers of parameters.  For very small numbers of configurations, the
optimizations give pathological results, especially when the more
flexible Jastrow factor is used.

Results obtained using reweighted variance minimization are also shown
in Fig.~\ref{VMC_E_v_Nc_neon}.  The reweighted variance-minimization
process was pathologically unstable for fewer than about $10^3$
configurations. For larger numbers of configurations the energies
obtained are in good agreement with the results of unreweighted
variance minimization.

\subsection{The distribution of configurations}

\begin{figure}
\begin{center}
\includegraphics*{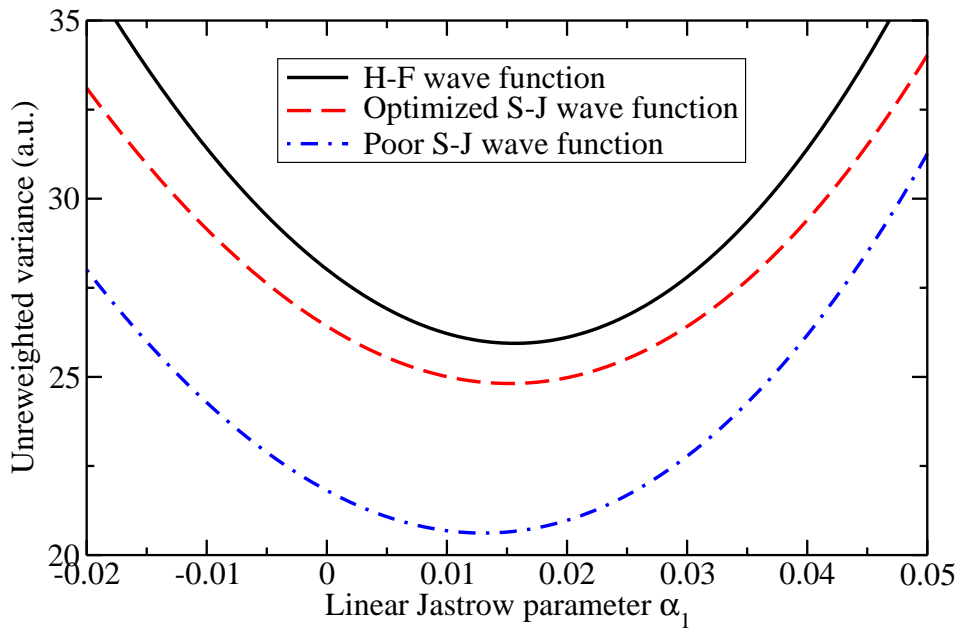}
\caption{The unreweighted variance for an all-electron neon atom
  plotted against the value of the linear Jastrow parameter $\alpha_1$
  for three different configuration distributions: the square of the
  Hartree-Fock wave function, the square of an optimized
  Slater-Jastrow wave function and the square of a poor Slater-Jastrow
  wave function.  $10^6$ configurations were used to calculate the
  unreweighted variance.  The Jastrow factor contained a total of 36
  linear parameters.  The Slater wave function contained Hartree-Fock
  orbitals.
  \label{fig:neon_distbn_choice}}
\end{center}
\end{figure}

The unreweighted variance for an all-electron neon atom is plotted
against a linear Jastrow parameter for three different configuration
distributions in Fig.~\ref{fig:neon_distbn_choice}.  The
configurations were distributed according to (i) the square of the
Hartree-Fock wave function, as is usually the case in the first cycle
of a variance-minimization calculation; (ii) the square of an
optimized Slater-Jastrow wave function, as is usually the case in the
second and subsequent cycles; and (iii) the square of a Slater-Jastrow
wave function in which the Jastrow factor was chosen to be poor.
Although the variance looks different in each case, the positions of
the minima coincide almost exactly for the Slater and optimized
Slater-Jastrow distributions. Even for the poor wave function, the
minimum of the variance is reasonably close to the more accurately
determined optimum.  This is consistent with our observation that, in
general, the only significant improvement to the quality of a Jastrow
factor occurs in the first cycle of a series of unreweighted
variance-minimization calculations: starting from the Hartree-Fock
wave function, the self-consistent solution is usually reached in the
first cycle.

\section{The flexibility of the Jastrow factor \label{section:flex}}

\begin{figure}
\begin{center}
\includegraphics*{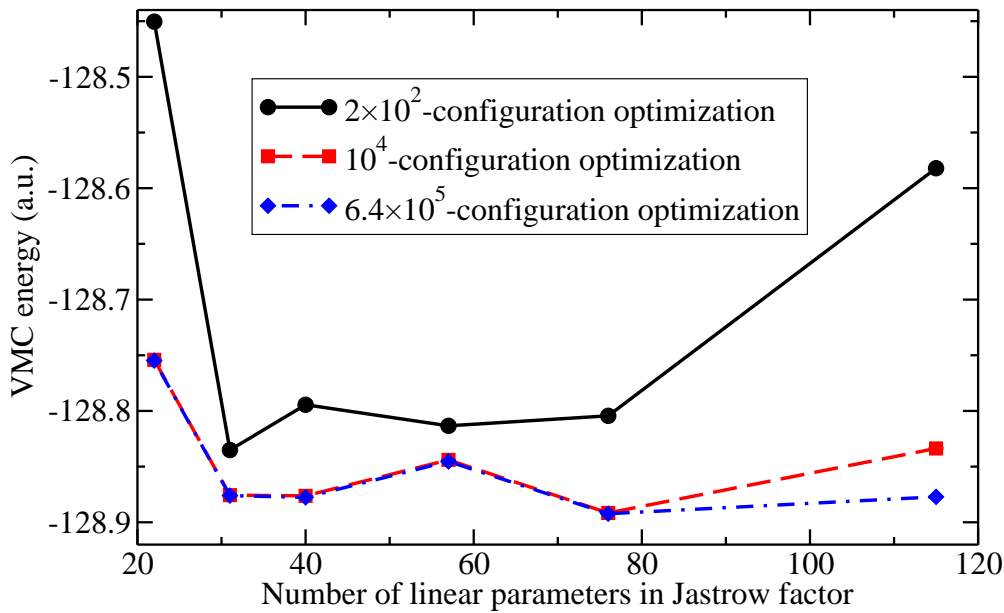}
\caption{The VMC energy of an all-electron neon atom against the
  number of parameters in the Jastrow factor.  Different numbers of
  configurations were used to carry out the unreweighted
  variance-minimization calculations.  Six optimization cycles were
  performed in order to guarantee self-consistency.  Very long VMC
  simulations were carried out using the optimized Jastrow factors in
  order to obtain the energies plotted in the graph.  The VMC error
  bars are smaller than the symbols.  The Slater wave function
  contained Hartree-Fock orbitals. \label{VMC_E_v_P_neon}}
\end{center}
\end{figure}

The VMC energy of neon is plotted against the number of linear
parameters used in the Jastrow factor in Fig.~\ref{VMC_E_v_P_neon}.
The results illustrate the futility of attempting to optimize too many
parameters.  The quality of the optimized wave function depends on the
number of configurations used to perform the optimization, especially
when the number of parameters in the wave function is either very
small or very large.  However, there would only appear to be an
advantage to be gained by using more than $10^4$ configurations when a
very large number of parameters are to be optimized.

It should be reemphasized that the problems which occur when large
numbers of parameters are optimized are caused by mismatches between
the minima of the unreweighted variance and the energy, and not by the
introduction of local minima into the variance landscape.

\section{Limiting of configuration weights}

It has been suggested that variance-minimization calculations are
disproportionately affected by ``outlying'' configurations, whose
energies deviate substantially from the mean energy.\cite{kent_1999}
In particular, the local energy diverges in the vicinity of the nodal
surface of the trial wave function, so configurations in this region
are especially problematic.  Such configurations are relatively rare
when the nodes are fixed, as is the case when only Jastrow parameters
are optimized, but the problem can be far more serious when
parameters that affect the nodal surface are optimized using a fixed
sampling of configuration space.

We have studied a smooth scheme for removing outlying configurations
from the optimization process.  Let us define the configuration
``effective weight'' to be
\begin{equation} W^\prime({\bf R}) = \frac{1}{2} \left[ 1 - \tanh \left(
  \frac{\left(E_L^{\{\alpha\}}({\bf R})-\bar{E}_u \right)^2-A^2
  \sigma^2_u}{B^2 \sigma^2_u} \right) \right], \end{equation} where
  $\bar{E}_u$ and $\sigma^2_u$ are the unreweighted energy and
  variance of the set of configurations.  $W^\prime({\bf R}) \approx
  1$ for configurations such that $E_L^{\{\alpha\}}({\bf R}) \approx
  \bar{E}_u$, but $W^\prime({\bf R}) \rightarrow 0$ for configurations
  whose local energies are far from the mean.  The parameter $A$ is
  the number of standard deviations of the energy beyond which
  configurations are excluded, while $B$ is the width of the region in
  which the effective weights fall off to zero (in terms of standard
  deviations of the energy).  We typically chose $A$ to lie between 2
  and 3 and $B$ to lie between $1/2$ and $1$.  The effective weights
  $W^\prime$ are used in place of the weights $W$ in
  Eq.~(\ref{eqn:rew_var_def}), and the reweighted variance
  $\sigma_w^2$ is minimized.

We have found that this weight-limiting scheme is capable of improving
the stability of Jastrow-factor optimization when very small numbers
of configurations are used.  However, the energies of the resulting
wave functions are not generally as good as the energies obtained
using the same forms of wave function optimized with an adequate
number of configurations.  For large numbers of configurations, the
limiting scheme has very little effect on the optimization of Jastrow
factors.  We conclude that the weight-limiting scheme is not of much
practical benefit when a Jastrow factor is to be optimized; however
the scheme has been found to be very useful when parameters that
affect the nodal surface are optimized.\cite{plr_comm}

Other limiting schemes have been devised to improve the stability of
the variance-minimization algorithms.  For example, it is possible to
combine the reweighted and unreweighted variance-minimization
algorithms by limiting the values that the weights $W$ can
take.\cite{filippi_1996}  Alternatively, the local energies themselves
can be limited.\cite{kent_1999}  The latter approach has been found to
be problematic, as it can result in spurious minima in the variance
corresponding to parameter sets for which a large number of local
energies are limited.

\section{Scaling of the variance-minimization methods
  \label{section:newopt_scaling}}

\subsection{CPU time required for the optimization phase}

Let $N$ be the total number of electrons in a system, $P$ be the
number of Jastrow parameters to be optimized, and $N_C$ be the number
of configurations used to calculate the variance.  Although the
Jastrow factor of Ref.~\cite{ndd_jastrow} is considered in this work,
the conclusions reached should be valid for most other forms of
Jastrow factor in current use.

The computational effort required to evaluate the unreweighted
variance using the accelerated method is independent of $N$ and $N_C$,
but scales as $O(P^4)$.  The time taken to compute the gradient of the
variance is also $O(P^4)$.  It may be assumed that the number of
optimization steps required is independent of $P$, $N$, and $N_C$.
The $O(P^4)$ scaling of the memory requirements of the accelerated
method limits the number of parameters that can be optimized in a
single calculation to between 100 and 200, depending on the available
memory.

The time taken to recompute the Jastrow factor and its derivatives
after all of the parameters have changed is generally $O(N)$ for
electron-nucleus and electron-electron-nucleus terms and $O(N^2)$ for
electron-electron terms.\cite{ndd_jastrow}  The CPU time required to
evaluate the variance (reweighted or unreweighted) using the standard
procedure therefore increases as $O(N^2)$.  The time taken to
calculate the Jastrow factor is, in general, $O(P)$, and hence the
time taken to calculate the variance using the standard method is also
$O(P)$.  Furthermore, each minimization step requires the gradient of
the variance with respect to the parameters, which has $P$ components
The time taken to perform each iteration is therefore $O(P^2)$.  The
CPU time for the standard method clearly scales as $O(N_C)$.

Putting this together, the CPU time for the optimization phase scales
as $O(P^4)$ for the accelerated method and $O(N^2P^2N_C)$ for the
standard method.  It should be noted that the time required by the
optimization phase in the accelerated scheme is completely negligible
in comparison with the time required by the VMC coefficient-gathering
phase, whereas the CPU time required by the optimization phase in the
standard method is usually rather greater than the CPU time required
by the VMC phase.

\subsection{The CPU time required for the gathering of the quartic
  coefficients in the accelerated scheme}

In the standard variance-minimization method, the CPU time required to
generate the set of configurations used to compute the variance does
not differ appreciably from the time taken to perform an ordinary VMC
simulation.  For the accelerated optimization method, however, the
time taken to compute the quartic expansion coefficients can be a
significant fraction of the total CPU time.

The gathering of the quartic coefficients can be divided into two
stages: (i) the evaluation of the Jastrow ``basis functions''
$f_i({\bf R})$ for each configuration ${\bf R}$ (see
Eq.~(\ref{eqn:linear_Jastrow_defn})), and (ii) the calculation of the
corresponding contributions to the $\bar{g}$ and $\bar{G}$
arrays. Stage (ii) scales as $O(P^4)$, but is independent of system
size.  By contrast, stage (i) scales as $O(P)$, because there are $P$
basis functions, but the scaling with system size is the same as that
of evaluating the Jastrow factor: roughly $O(N^2)$.

The CPU time for an ordinary VMC calculation is generally determined
by the time taken to evaluate the orbitals in the Slater wave
function. The computational effort required to carry out a fixed
number of configuration moves grows as $O(N^2)$ if extended orbitals
represented in a localized basis are used.  The use of localized
orbitals can improve this scaling to
$O(N)$.\cite{williamson_lin_scaling} In principle the time taken for
stage (i) of the coefficient gathering will take up an increasingly
large fraction of the CPU time, but in practice the prefactor is so
small that the time required is negligible even for the largest
systems that we have studied.  The time taken for stage (ii) can be
the largest contribution to the CPU time for VMC simulations of small
molecules, but the effort required is independent of system size, and
so, overall, the coefficient-gathering phase of the accelerated scheme
is more efficient for large systems than small systems.

\section{Efficiency of the accelerated optimization method
  \label{section:newopt_timing}}

Timing results for the optimization of the linear Jastrow parameters
for an H$_2$O molecule (10 electrons) and a C$_{26}$H$_{32}$ molecule
(136 electrons) are shown in Tables \ref{table:h2o_timing} and
\ref{table:c26h32_timing} respectively.  The calculations are fairly
typical in terms of the number of parameters and number of
configurations.  In both cases the use of the accelerated optimization
scheme essentially eliminates the cost of the optimization phase.  In
the standard method the cost of the optimization phase exceeds that of
the VMC configuration-generation phase by an order of magnitude for
H$_2$O and by a less significant proportion for C$_{26}$H$_{32}$.  The
cost of the VMC phase in the accelerated scheme is increased
substantially for H$_2$O although, overall, it is still much faster to
use the accelerated scheme.  For C$_{26}$H$_{32}$ the increase in the
CPU time for configuration generation is negligible.  Overall, the
accelerated optimization scheme is 4.5 times faster for H$_2$O and 2.3
times faster for C$_{26}$H$_{32}$.

\begin{table}
\begin{center}
\begin{tabular}{lcr@{.}l}
\hline \hline

Method & Stage & \multicolumn{2}{c}{CPU time (s)} \\

\hline

         & VMC   & $5669$&$43$ \\

Standard & Opt.  & $58740$&$65$  \\

         & Total & \hspace{0.3cm} $64410$&$08$  \\

\hline

         & VMC   & $14378$&$90$  \\

Accel.   & Opt.  & $39$&$69$       \\

         & Total & $14418$&$59$  \\

\hline \hline
\end{tabular}
\caption{Timing results for ten cycles of a $6 \times
  10^4$-configuration unreweighted variance minimization of a
  38-linear-parameter Jastrow factor for an all-electron H$_2$O
  molecule.  The system contains a total of 10 electrons.  The Slater
  wave function contained Hartree-Fock orbitals.  The runs were
  carried out on a 1.7\,GHz Pentium processor in a Sony Vaio laptop.
  \label{table:h2o_timing}}
\end{center}
\end{table}

\begin{table}
\begin{center}
\begin{tabular}{lcr@{.}l}
\hline \hline

Method & Stage & \multicolumn{2}{c}{CPU time (s)} \\

\hline

         & VMC   & $6526$&$77$ \\

Standard & Opt.  & $9323$&$53$   \\

         & Total & \hspace{0.3cm} $15850$&$30$  \\

\hline

         & VMC   & $6828$&$47$   \\

Accel.   & Opt.  & $28$&$14$       \\

         & Total & $6856$&$61$   \\

\hline \hline
\end{tabular}
\caption{Timing results for four cycles of a $1.6 \times
  10^4$-configuration unreweighted variance minimization of a
  12-linear-parameter Jastrow factor for a C$_{26}$H$_{32}$ molecule
  with Troullier-Martins carbon and hydrogen pseudopotentials.  The
  system contains a total of 136 electrons.  The Slater wave function
  contained DFT-PBE orbitals.  The runs were carried out on a cluster
  of eight 2.1\,GHz Opteron processors.
  \label{table:c26h32_timing}}
\end{center}
\end{table}

The actual time taken to compute the variance in the accelerated
scheme is minute:  On a 2.7\,GHz Pentium 4 processor, it takes an
average of $83.6$\,$\mu$s to compute the variance with 25 parameters,
while it takes $13.98$\,ms to compute the variance with 100 parameters.

\section{Conclusions \label{section:newopt_conclusions}}

We have introduced a new scheme for evaluating the unreweighted
variance of the VMC energy, which greatly accelerates the optimization
of parameters that occur in a linear fashion in the exponent of a
Jastrow factor.  This scheme is very efficient because it uses the
property that the unreweighted variance is a quartic function of such
parameters. We studied a wide range of systems and found that the
unreweighted variance almost invariably has a single minimum in the
space of the linear parameters.  The only exceptions to this that we
could find occurred when the configuration space was very poorly
sampled.  For other wave-function parameters, however, the
unreweighted variance often has more than one minimum.

It is easy to use very large numbers of configurations to perform
optimizations using our accelerated scheme.  We have investigated the
effect of varying the number of configurations on the wave-function
quality, and we have found that there is, in general, no significant
benefit to be obtained from using more than about $10^4$
configurations when optimizing linear Jastrow  parameters.

We have considered various wave-function optimization schemes using
correlated-sampling approaches for minimizing the energy and the
variance of the energy.  Reweighted energy and variance minimization
using correlated sampling suffer from numerical instabilities due to
fluctuations in the values of the weights, which are severe for large
systems. The unreweighted energy always has a stationary point at the
wave function used to generate the configuration set, and for
parameters which occur linearly in the Jastrow factor this stationary
point is the global maximum in the energy.  The unreweighted energy is
therefore not a suitable cost function for wave-function optimization.
The minima of the variance, the unreweighted variance (iterated to
self-consistency), and the energy are generally distinct.  In various
model systems that we have studied, the self-consistent minimum in the
unreweighted variance always gave lower energies than the minimum in
the reweighted variance.

\section{Acknowledgments}

Financial support was provided by the Engineering and Physical
Sciences Research Council (EPSRC), UK\@.  Computing resources have
been provided by the Cambridge-Cranfield High Performance Computing
Facility.

\appendix

\section{Constructing the quartic polynomial corresponding to a line
  in parameter space \label{section:line_min_details}}

Consider the expression for the quartic unreweighted variance as a
function of the linear parameters
(Eq.~(\ref{eqn:lsq_in_terms_of_Gamma})), and consider a line in
parameter space
\begin{equation} \mbox{\boldmath $\alpha$}(t) = {\bf A}+{\bf B}t,
\end{equation}
where $\mbox{\boldmath $\alpha$}=(\alpha_1,\ldots,\alpha_P)$ and ${\bf
A}$ and ${\bf B}$ are constant vectors.  The unreweighted variance
along the line is given by
\begin{equation}
\sigma_u^2(t)=\frac{N_C}{N_C-1} \left( \Omega_4 t^4+\Omega_3
t^3+\Omega_2 t^2 +\Omega_1 t + \Omega_0 \right) ,
\end{equation}
where
\begin{eqnarray}
\Omega_4 & = & \sum_{i=1}^P B_i \sum_{j=i}^P B_j \sum_{k=j}^P B_k
\sum_{l=k}^P B_l \Gamma_{ijkl}^{(4)} \label{eqn:omega4_defn} \\
\Omega_3 & = & \sum_{i=1}^P A_i \sum_{j=i}^P B_j \sum_{k=j}^P B_k
\sum_{l=k}^P B_l \Gamma_{ijkl}^{(4)} + \sum_{i=1}^P B_i \sum_{j=i}^P
A_j \sum_{k=j}^P B_k \sum_{l=k}^P B_l \Gamma_{ijkl}^{(4)} \nonumber \\
& & {} + \sum_{i=1}^P B_i \sum_{j=i}^P B_j \sum_{k=j}^P A_k
\sum_{l=k}^P B_l \Gamma_{ijkl}^{(4)} + \sum_{i=1}^P B_i \sum_{j=i}^P
B_j \sum_{k=j}^P B_k \left( \Gamma_{ijk}^{(3)} + \sum_{l=k}^P A_l
\Gamma_{ijkl}^{(4)} \right) \\ \Omega_2 & = & \sum_{i=1}^P A_i
\sum_{j=i}^P A_j \sum_{k=j}^P B_k \sum_{l=k}^P B_l \Gamma_{ijkl}^{(4)}
+ \sum_{i=1}^P A_i \sum_{j=i}^P B_j \sum_{k=j}^P A_k \sum_{l=k}^P B_l
\Gamma_{ijkl}^{(4)} \nonumber \\ & & {} + \sum_{i=1}^P A_i
\sum_{j=i}^P B_j \sum_{k=j}^P B_k \left( \Gamma_{ijk}^{(3)} +
\sum_{l=k}^P A_l \Gamma_{ijkl}^{(4)} \right) + \sum_{i=1}^P B_i
\sum_{j=i}^P A_j \sum_{k=j}^P A_k \sum_{l=k}^P B_l \Gamma_{ijkl}^{(4)}
\nonumber \\ & & {} + \sum_{i=1}^P B_i \sum_{j=i}^P A_j \sum_{k=j}^P
B_k \left( \Gamma_{ijk}^{(3)} + \sum_{l=k}^P A_l \Gamma_{ijkl}^{(4)}
\right) \nonumber \\ & & {} + \sum_{i=1}^P B_i \sum_{j=i}^P B_j \left(
\Gamma_{ij}^{(2)} + \sum_{k=j}^P A_k \left( \Gamma_{ijk}^{(3)} +
\sum_{l=k}^P A_l \Gamma_{ijkl}^{(4)} \right) \right) \\ \Omega_1 & = &
\sum_{i=1}^P B_i \left( \Gamma_i^{(1)} + \sum_{j=i}^P A_j \left(
\Gamma_{ij}^{(2)} + \sum_{k=j}^P A_k \left( \Gamma_{ijk}^{(3)} +
\sum_{l=k}^P A_l \Gamma_{ijkl}^{(4)} \right) \right) \right) \nonumber
\\ & & {} + \sum_{i=1}^P A_i \sum_{j=i}^P B_j \left( \Gamma_{ij}^{(2)}
+ \sum_{k=j}^P A_k \left( \Gamma_{ijk}^{(3)} + \sum_{l=k}^P A_l
\Gamma_{ijkl}^{(4)}\right) \right) \nonumber \\ & & {} + \sum_{i=1}^P
A_i \sum_{j=i}^P A_j \sum_{k=j}^P B_k \left( \Gamma_{ijk}^{(3)} +
\sum_{l=k}^P A_l \Gamma_{ijkl}^{(4)}\right) + \sum_{i=1}^P A_i
\sum_{j=i}^P A_j \sum_{k=j}^P A_k \sum_{l=k}^P B_l \Gamma_{ijkl}^{(4)}
\\ \Omega_0 & = & \Gamma^{(0)} + \sum_{i=1}^P A_i \left(
\Gamma_i^{(1)} + \sum_{j=i}^P A_j \left( \Gamma_{ij}^{(2)} +
\sum_{k=j}^P A_k \left( \Gamma_{ijk}^{(3)} + \sum_{l=k}^P A_l
\Gamma_{ijkl}^{(4)} \right) \right) \right).
\label{eqn:omega0_defn}
\end{eqnarray}
All the terms that appear in
Eqs.~(\ref{eqn:omega4_defn})--(\ref{eqn:omega0_defn}) can be evaluated
within a single loop over $i$, $j$, $k$, and $l$.


\begin{thebibliography}{9}

\bibitem{foulkes_2001} W.~M.~C.~Foulkes, L.~Mitas, R.~J.~Needs, and
G.~Rajagopal, Rev.\ Mod.\ Phys.\ \textbf{73}, 33 (2001).

\bibitem{umrigar_1988a} C.~J.~Umrigar, K.~G.~Wilson, and
J.~W.~Wilkins, Phys.\ Rev.\ Lett.\ \textbf{60}, 1719 (1988).

\bibitem{kent_1999} P.~R.~C.~Kent, R.~J.~Needs, and G.~Rajagopal,
Phys.\ Rev.\ B \textbf{59}, 12344 (1999).

\bibitem{casino} R.~J.~Needs, M.~D.~Towler, N.~D.~Drummond, and
P.~R.~C.~Kent, \textsc{casino} version 1.7 User Manual, University of
Cambridge, Cambridge (2003).

\bibitem{footnote_unreweighted_energy} In Ref.~\cite{kent_1999} it was
stated that the unreweighted energy could have a maximum, minimum, or
saddle point at the parameter values $\{ \alpha_0 \}$, but this is
incorrect; we have proved that it is always a maximum.

\bibitem{ndd_jastrow} N.~D.~Drummond, M.~D.~Towler, and R.~J.~Needs,
  Phys.\ Rev.\ B \textbf{70}, 235119 (2004).

\bibitem{press_F77} W.~H.~Press, S.~A.~Teukolsky, W.~T.~Vetterling,
and B.~P.~Flannery, \textit{Numerical Recipes in Fortran 77} (2nd
ed.), Cambridge University Press (1992).

\bibitem{lee_thesis} Y.~Lee, PhD Thesis, University of Cambridge,
  Cambridge (2002).

\bibitem{plr_comm} P.~Lopez Rios, Personal communication (2005).

\bibitem{filippi_1996} C.~Filippi and C.~J.~Umrigar, J.\ Chem.\ Phys.\
  \textbf{105}, 213 (1996).

\bibitem{williamson_lin_scaling} A.~J.~Williamson, R.~Q.~Hood, and
  J.~C.~Grossman, Phys.\ Rev.\ Lett.\ \textbf{87}, 246406 (2001).

\end{thebibliography}
\end{document}